\documentstyle[11pt,a4,cite,epsf]{article}
\newcommand{\lesssim}{ {\
\lower-1.2pt\vbox{\hbox{\rlap{$<$}\lower5pt\vbox{\hbox{$\sim$}}}}\ } }
\newcommand{\gtrsim}{ {\
\lower-1.2pt\vbox{\hbox{\rlap{$>$}\lower5pt\vbox{\hbox{$\sim$}}}}\ } }

\newcommand{\be}{\begin{equation}}
\newcommand{\ee}{\end{equation}}
\newcommand{\bea}{\begin{eqnarray}}
\newcommand{\eea}{\end{eqnarray}}
\newcommand{\noi}{\noindent}
\newcommand{\nn}{\nonumber}

\newcommand{\cL}{{\cal L}}
\newcommand{\cO}{{\cal O}}
\newcommand{\cH}{{\cal H}}

\newcommand{\Imm}{\mbox{\rm Im}}

\newcommand{\tr}{\mbox{\rm tr}}

\newcommand{\annd}{\mbox{\rm and}}
\newcommand{\foor}{\mbox{\rm for}}

\newcommand{\al}{\alpha}
\newcommand{\als}{\alpha_{\mbox{\rm {\scriptsize s}}}}

\newcommand{\SM}{\mbox{\rm {\tiny SM}}}
\newcommand{\NC}{\mbox{\rm {\tiny NC}}}

\newcommand{\QCD}{QCD$(\infty)\;$}

\newcommand{\eff}{\mbox{\rm{\tiny eff}}}

\newcommand{\EM}{\mbox{\tiny\rm  EM}}

\newcommand{\qcd}{\mbox{\rm{\tiny QCD}}}
\newcommand{\diag}{\mbox{\rm diag}}

\newcommand{\ra}{\rightarrow}

\input epsf

\begin{document}
\begin{titlepage}

\begin{flushright} CPT-98/P.3701\\ UAB-FT-453\\ \today
\end{flushright}
\vspace*{1.5cm}
\begin{center} {\Large \bf The Electroweak $\pi^{+}-\pi^{0}$ Mass
Difference\\[0.1cm] 
and \\[0.2cm]
Weak Matrix Elements in the 
1/${N_c}$ Expansion}\\[3.0cm]  {\large {\bf Marc Knecht}$^{a}$, 
{\bf Santiago Peris}$^b$ and {\bf Eduardo de Rafael}$^a$}\\[1cm]

$^a$  Centre  de Physique Th{\'e}orique\\
       CNRS-Luminy, Case 907\\
    F-13288 Marseille Cedex 9, France\\[0.5cm]
$^b$ Grup de F{\'\i}sica Te{\`o}rica and IFAE\\ Universitat Aut{\`o}noma de
Barcelona, 08193 Barcelona, Spain.\\

\end{center}

\vspace*{1.0cm}

\begin{abstract}

The $\pi^{+}-\pi^{0}$ mass difference generated by the electroweak
interactions of the Standard Model is expressed, to lowest order in the chiral
expansion and to leading order in $\alpha$, in terms of an integral
involving the
same correlation function
$\Pi_{LR}(Q^2)$  which governs the well known electromagnetic $\pi^{+}-\pi^{0}$
mass difference. We calculate this contribution within the framework of QCD
in the
limit of a large number of colours $N_c$. We show how this calculation, which
requires non--trivial contributions from next--to--leading terms in the
$1/N_c$ expansion, provides an excellent theoretical laboratory for studying
issues of long-- and short--distance matching in calculations of weak matrix
elements of four--quark operators. The electroweak $\pi^{+}-\pi^{0}$ mass
difference
turns out to be a physical observable which is under good analytical control and
which should, therefore, be an excellent testground of numerical evaluations in
lattice--QCD  and of model calculations in general.
\end{abstract}
\end{titlepage}

\section{Introduction}
\label{sec:Introduction}

In the chiral limit where the light quark masses are set to zero, the
low--energy
effective Lagrangian of QCD in the presence of electromagnetic interactions
has, to
lowest order in $\al={e^2}/{4\pi}\simeq 1/137.03...$, an interaction term
without
derivatives of order
$\cO (p^0)$ in the chiral expansion~\cite{EGPR89}
\be\label{eq:effem}
\cL_{\mbox{\scriptsize\rm eff}}=\cdots +e^2 C\,\tr
\left(Q_{R}UQ_{L}U^{\dagger}\right)\,.
\ee Here, $U$ is the matrix field which collects the octet of pseudoscalar
Goldstone fields and $Q_{R}=Q_{L}=\mbox{\rm diag}[2/3, -1/3, -1/3]$ the
right-- and
left--handed charges associated with the electromagnetic couplings of the light
quarks. Upon expanding $U$ in powers of the pseudoscalar fields there appear
quadratic terms like
\be\label{eq:effemex}
\cL_{\mbox{\scriptsize\rm eff}}=\cdots -2e^2 C\frac{1}{f_{\pi}^2} (\pi^+
\pi^- + K^+ K^-) + \cdots ,
\ee showing that, in the presence of electromagnetic interactions, the
charged pion
and kaon fields become massive in agreement with the old  current algebra result
obtained by Dashen~\cite{Dashen69},
\be (m_{K^+}^{2}-m_{K^0}^{2})\vert_{\mbox{\scriptsize\rm
EM}}=(m_{\pi^+}^{2}-m_{\pi^0}^{2})\vert_{\mbox{\scriptsize\rm EM}}=
\frac{2e^2 C}{f_{\pi}^2}+ \cdots.
\ee In fact, the main contribution to the physical
$\pi^{+}-\pi^{0}$ mass difference {\em is} of electromagnetic origin
because quark
masses do not contribute significantly to the
$\pi^{+}-\pi^{0}$ mass splitting~\cite{GL82}.

The effective term in eq.~(\ref{eq:effem}) results from the integration of a
virtual photon in the presence of the strong interactions in the chiral
limit. It
is well known~\cite{Lowetal67,BdeR91,Bachir97,KU97} that the corresponding
coupling
constant
$C$ is given by an integral of a correlation function
$\Pi_{LR}(Q^2)$ which is the invariant amplitude of the two--point function
\be\label{eq:lrtpf}
\Pi_{LR}^{\mu\nu}(q)=2i\int d^4 x\,e^{iq\cdot x}\langle 0\vert
\mbox{\rm T}\left(L^{\mu}(x)R^{\nu}(0)^{\dagger}
\right)\vert 0\rangle\,,
\ee with currents
\be L^{\mu}=\bar{d}(x)\gamma^{\mu}\frac{1}{2}(1-\gamma_{5})u(x)
\quad \annd \quad R^{\mu}=\bar{d}(x)\gamma^{\mu}\frac{1}{2}(1+\gamma_{5})u(x)\,.
\ee In the chiral limit
\be\label{eq:lritpf}
\Pi_{LR}^{\mu\nu}(Q^2)=(q^{\mu}q^{\nu}-g^{\mu\nu}q^2)\Pi_{LR}(Q^2)\,,
\ee and
\be\label{eq:piem}
C=\frac{-1}{8\pi^2}\,\frac{3}{4}\,
\int_0^\infty dQ^2\,Q^2\Pi_{LR}(Q^2)\,,
\ee
with $Q^2$ the euclidean momentum squared of the virtual photon.
   The function $\Pi_{LR}(Q^2)$ is, for all values of $Q^2$, an order
parameter of
spontaneous chiral symmetry breakdown (S$\chi$SB)  and thus, so is the
low--energy
constant $C$.  This means that the correlator in eq.~(\ref{eq:lrtpf}) has a
smooth
behaviour at short distances, so that the integral in eq.~(\ref{eq:piem})
converges
in the ultraviolet region~:
\be\label{eq:wsrs}
\lim_{Q^2\ra\infty} Q^4 \Pi_{LR}(Q^2)\ra 0.
\ee This behaviour also entails the two Weinberg sum rules~\cite{We67} in the
chiral limit. It has furthermore been shown~\cite{W83, CLT95} that
\be\label{eq:witten} -Q^2\Pi_{LR}(Q^2)\ge 0 \qquad\foor\;  \quad 0\le
Q^2\le\infty\,,
\ee which in particular ensures the positivity of the integral in
eq.~(\ref{eq:piem}) and thus the stability of the QCD vacuum with respect
to  small
perturbations induced by electromagnetic interactions.

In this letter we shall first show that the same correlation function
$\Pi_{LR}(Q^2)$ governs the $\pi^{+}-\pi^{0}$ mass difference in the presence of
the electroweak interactions of the Standard Model. This observation and the
calculation which follows thereof are of course of phenomenological
interest {\it
per se}; but their real relevance lies in the fact that, as we shall show below,
they provide an excellent theoretical laboratory to study issues of {\it
long-- and
short--distance matching} in calculations of weak matrix elements of four--quark
operators. We also think that this example, which is under good analytical
control,
provides an excellent testground for numerical evaluations in 
lattice--QCD and for 
model calculations in general.


\section{The Electroweak $\pi^{+}-\pi^{0}$ Mass Difference}
\label{sec:Elweakmass}

In the presence of the electromagnetic and weak neutral currents of the
Electroweak
Model, the QCD lagrangian $\cL_{\qcd}$ becomes
\be
\cL_{\qcd}\ra \cL_{\qcd}+
\bar{q}_{R}\gamma^{\mu}r_{\mu}q_{R} +\bar{q}_{L}\gamma^{\mu}l_{\mu}q_{L}\,,
\ee where
\bea\label{eq:external} l_{\mu} &= & eQ_{L}[A_{\mu}(x)-\tan\theta_{W}
Z_{\mu}(x)]+
\frac{e}{2\sin\theta_{W}\cos\theta_{W}}T_{3}Z_{\mu}(x)\,,\\ r_{\mu} &= &
eQ_{R}[A_{\mu}(x)-\tan\theta_{W} Z_{\mu}(x)]\,,
\eea and
\be Q_{L}=Q_{R}=\frac{1}{3}\diag (2,-1,-1)\,,\qquad T_{3}=\diag(1,-1,-1)\,.
\ee There are then two possible terms of order $\cO(p^0)$ in the low--energy
effective lagrangian generated by the virtual
$Z$ integration~\footnote{The charged $W$--field only couples to left currents
and therefore cannot contribute to an $\cO(p^0)$ self--energy.}:
\be
+e^2 \tan^2 \theta_{W} C_{Z}\tr (Q_{R}UQ_{L}U^{\dagger})\quad\annd\quad -e^2
\frac{1}{2\cos^2
\theta_{W}} C_{Z}\tr (Q_{R}UT_{3}U^{\dagger})\,,
\ee
where
\be\label{eq:CZ}
C_{Z}=\frac{1}{8\pi^2}\,\frac{3}{4}\,
\int_0^\infty dQ^2\,\frac{Q^2}{Q^2 +M_{Z}^2}\,Q^2\Pi_{LR}(Q^2)\,.
\ee
These couplings give rise to physical mass terms
\be
+e^2 \tan^2 \theta_{W} C_{Z}\tr (Q_{R}UQ_{L}U^{\dagger})=-2e^2
\frac{\sin^2 \theta_{W}}{\cos^2 \theta_{W}}
C_{Z}\frac{1}{f_{\pi}^2}\left(\pi^{+}\pi^{-}+ K^{+}K^{-}\right)+\cdots\,,
\ee
and
\be
-e^2 \frac{1}{2\cos^2 \theta_{W}} C_{Z}\tr (Q_{R}UT_{3}U^{\dagger})= 2e^2
\frac{1}{\cos^2 \theta_{W}} C_{Z}\frac{1}{f_{\pi}^2}
\left(\pi^{+}\pi^{-}+K^{+}K^{-}\right)+\cdots\,.
\ee
Collecting together these  weak neutral current contributions with the
electromagnetic contribution in eq.~(\ref{eq:effemex}), we arrive at the quite
remarkable result that, to lowest order in $e^2$ and in the chiral limit,
\be\label{eq:ewmass}
(m_{K^+}^2-m_{K^0}^2)\vert_{\SM}=
(m_{\pi^+}^2-m_{\pi^0}^2)\vert_{\SM}=
\frac{2e^2 C_{\SM}}{f_{\pi}^2}\,,
\ee with
\be\label{eq:CSM}
C_{\SM}=\frac{1}{8\pi^2}\frac{3}{4}\int_{0}^{\infty} dQ^2
\left(1-\frac{Q^2}{Q^2 +M_{Z}^2}\right) \left(-Q^2
\Pi_{LR}(Q^2)\right)\,.
\ee
In the limit where $M_{Z}\ra\infty$ we recover the usual expression of Low {\it
et al.}~\cite{Lowetal67}, while there is no induced mass difference in the limit
$M_{Z}\ra 0$ where the $SU(2)\times U(1)$ gauge symmetry is unbroken. This can
readily be seen by rewriting the external fields in eq.~(\ref{eq:external}) in
terms of the usual
$W_{\mu}^{3}$ and $B_{\mu}$ gauge fields:
\bea\label{eq:externalBW}
l_{\mu} &= &
\frac{e}{\cos\theta_{W}}\left[Q_{L}-\frac{1}{2}T_{3}\right]B_{\mu}(x) +
\frac{e}{2\sin\theta_{W}}T_{3}W_{\mu}^{3}(x)\,,\\ r_{\mu} &= &
\frac{e}{\cos\theta_{W}}Q_{R}B_{\mu}(x)\,.
\eea Neither the $W_{\mu}^{3}$ field, which has a pure left--handed coupling to
quarks, nor the
$B_{\mu}$ field, which has a left component proportional to the unit
matrix, can
generate an effective $\cO(p^0)$ interaction of Goldstone fields. When the
$SU(2)\times U(1)$ gauge symmetry is broken down to $U(1)_{\EM}$,
eqs.~(\ref{eq:ewmass}) and (\ref{eq:CSM}) show that the net effect of the
massive
physical
$Z$ is to reduce just a little bit the $\pi^{+}$ mass induced by
electromagnetism
alone.

It is possible to make a phenomenological evaluation of the integral in
eq.~(\ref{eq:CSM})  using the fact that the function $\Pi_{LR}(Q^2)$ obeys an
unsubtracted dispersion relation
\be\label{eq:disp}
\Pi_{LR}(Q^2)=\int_{0}^{\infty}dt\frac{1}{t+Q^2}\left[\frac{1}{\pi}\Imm
\Pi_{V}(t)-\frac{1}{\pi}\Imm\Pi_{A}(t)\right]\,,
\ee
with $\frac{1}{\pi}\Imm
\Pi_{V}(t)$ and $\frac{1}{\pi}\Imm
\Pi_{A}(t)$  the physical vector and axial--vector spectral functions which are
measured in hadronic $e^{+} e^{-}$ annihilations and in hadronic
$\tau$ decays.
The physical spectral functions, however, are not quite the ``chiral limit''
spectral functions which should be inserted in eq.~(\ref{eq:CSM}). It is
possible to do the appropriate corrections, as discussed e.g. in
ref.~\cite{DGHS98}, but for the purposes of this letter we wish to follow a more
theoretical procedure. We propose to evaluate the integral in eq.~(\ref{eq:CSM})
within the framework of QCD in the limit of a large number of colours
$N_c$~\cite{'tH74,RV77,W79}, which we shall denote by \QCD for short.

The spectral function associated with
$\Pi_{LR}(Q^2)$ in \QCD consists of the difference of an infinite number of
narrow
vector states and an infinite number of narrow axial--vector states,
together with
the Goldstone pion pole:
\be
\frac{1}{\pi}\Imm\Pi_{LR}(t) =\sum_{V}f_{V}^2 M_{V}^2\delta(t-M_{V}^2)
-\sum_{A}f_{A}^2 M_{A}^2\delta(t-M_{A}^2)-f_{\pi}^2\delta(t)\,.
\ee Since $\Pi_{LR}(Q^2)$ obeys the dispersion relation~(\ref{eq:disp}), we find
that
\be\label{eq:LRN1}
-Q^2\Pi_{LR}(Q^2)=f_{\pi}^2+\sum_{A}f_{A}^2
M_{A}^2\frac{Q^2}{M_{A}^2+Q^2} -\sum_{V}f_{V}^2
M_{V}^2\frac{Q^2}{M_{V}^2+Q^2}\,.
\ee
Furthermore, the two Weinberg sum rules that follow from eq.~(\ref{eq:wsrs})
constrain the couplings and masses of the narrow states as follows:
\be\label{eq:weinbergsrs}
\sum_{V}f_{V}^2 M_{V}^2-\sum_{A}f_{A}^2 M_{A}^2=f_{\pi}^2
\quad
\annd\quad
\sum_{V}f_{V}^2 M_{V}^4-\sum_{A}f_{A}^2 M_{A}^4=0\,,
\ee
ensuring the convergence of the integral in eq.~(\ref{eq:CSM}) in \QCD, with the
result
\be\label{eq:CSMexact}
C_{\SM}=\frac{-3}{32\pi^2}\left[\sum_{A}\frac{M_{Z}^2}{M_{Z}^2-M_{A}^2}f_{A}^2
M_{A}^4
\log\frac{M_{Z}^2}{M_{A}^2}-\sum_{V}\frac{M_{Z}^2}{M_{Z}^2-M_{V}^2}f_{V}^2
M_{V}^4
\log\frac{M_{Z}^2}{M_{V}^2}\right]\,.
\ee

The original evaluation by Low {\it et al.}~\cite{Lowetal67} of the integral in
eq.~(\ref{eq:piem})  was made by considering only the phenomenological
contributions from the pion pole and the lowest vector and axial--vector
states in
the narrow width approximation. This approximation, which reproduces very
well the
experimental determination of the physical
$\pi^{+}-\pi^{0}$ mass difference, can nowadays be viewed as the
approximation to
\QCD which consists in restricting the part of the hadronic spectrum which is
responsible for S$\chi$SB in the channels with the $J^{P}$ quantum numbers
$1^{-}$ and $1^{+}$ to their lowest energy states. The rest of the infinite
number
of narrow states are then treated as dual to their corresponding
\QCD perturbative continuum, and therefore do not contribute to order
parameters of
S$\chi$SB. An explicit formulation in terms of an effective Lagrangian of this
lowest meson dominance (LMD) approximation to \QCD, with inclusion of the
$0^-$ and
$0^{+}$ channels as well,  has been recently
discussed in ref.~\cite{PPdeR98}. In the LMD approximation to
\QCD, the integral in eq.~(\ref{eq:CSM}) gives the result
\be\label{eq:CSMLMD} C_{\SM}\simeq\frac{3}{4}\frac{f_{\pi}^2}{8\pi^2}
M_{V}^2\left[
\frac{M_{A}^2}{M_{A}^2-M_{V}^2}\log\frac{M_{A}^2}{M_{V}^2}-
\frac{M_{A}^2}{M_{Z}^2}\left(\log\frac{M_{Z}^2}{M_{V}^2}-
\frac{M_{A}^2}{M_{A}^2-M_{V}^2}\log\frac{M_{A}^2}{M_{V}^2}
\right)\right]\,,
\ee
where we have expanded in powers of $1/M_{Z}^2$ and retained only the leading
term. The first term in the r.h.s. of eq.~(\ref{eq:CSMLMD}) is the contribution
from the virtual photon integration. The
$Z$--induced contribution consists of a ``large''
$\log M_{Z}^2$ term and a ``constant'' term. Overall, it represents a
correction of 
0.097\% \footnote{For the numerical evaluation, we have taken $M_V$ and
$M_A$ as the $\rho(770)$ and the $a_1(1260)$ masses, respectively. Our
normalization convention corresponds to $f_\pi = 92.4$ MeV.} 
to the lowest order
electromagnetic contribution. The effect, as expected, is very small but
nevertheless larger than the present experimental accuracy in the
determination of
the $\pi^{+}-\pi^{0}$ mass difference. Although this observation may bear some
interest by itself, we shall not pursue it any further here. We rather
dedicate the rest of this note to several theoretically interesting aspects
of this
calculation.

\section {$\Pi_{LR}(Q^2)$ and the Operator Product Expansion}
\label{sec:OPE}

As recently discussed in ref.~\cite{KdeR97}, besides
the two Weinberg sum rules in eqs.~(\ref{eq:weinbergsrs}), there are further
constraints between masses and couplings of resonances on the one hand and the
local order parameters which govern the operator product expansion (OPE) of the
$\Pi_{LR}(Q^2)$ function on the other. In particular, it has been
shown~\cite{KdeR97} that in the large--$N_c$ limit, the leading $d=6$ order
parameter is given by the expression
\bea\label{eq:phi3}
\lim_{Q^2\ra\infty}Q^6\Pi_{LR}(Q^2) & = &
-4\pi^2\left(\frac{\alpha_s}{\pi}+\cO(\alpha_s^2)\right)
\langle\bar{\psi}\psi\rangle^2 \nonumber\\ & = & \sum_{V}
f_{V}^2 M_{V}^6-\sum_{A} f_{A}^2 M_{A}^6\,,
\eea where $\langle\bar{\psi}\psi\rangle$ is the usual single flavour quark
bilinear condensate ($\psi=u,d,$ or $s$) in the chiral limit, and the Wilson
coefficient is the result of a lowest order calculation~\cite{SVZ79} in
powers of
$\alpha_s$. We shall show that it is precisely this
$d=6$ term which controls the $\log M_{Z}^2$ contribution to the
$Z$--induced part of the integral $C_{Z}$ in eq.~(\ref{eq:CZ}). Indeed,
following a Wilsonian approach, we can
split this integral into a low--energy region $0\le Q^2 \le
\mu^2 \ll M_Z^2$ where the $Z$ field is integrated out, {\it i.e.} where we approximate
${1}/({Q^2 +M_{Z}^2})$ by
${1}/{M_{Z}^2}$, and a high--energy region $\mu^2\le Q^2\le\infty$ where the
$Z$--propagator is fully kept but the
$\Pi_{LR}(Q^2)$ function is approximated by the leading $d=6$ term in
eq~(\ref{eq:phi3}). These approximations are equivalent to neglecting
higher order
corrections in $1/M_Z^2$ and in $1/\mu^2$. We then obtain
\bea\label{eq:OPEsplit} C_{Z} & = &
\frac{3}{32\pi^2}\frac{1}{M_{Z}^2}\left[\sum_{A}f_{A}^2 M_{A}^6
\log\frac{M_{A}^2}{\mu^2}-\sum_{V}f_{V}^2 M_{V}^6
\log\frac{M_{V}^2}{\mu^2}\right] \nn \\
 &  & \frac{-3}{32\pi^2}\frac{1}{M_{Z}^2}\left[\sum_{A}f_{A}^2 M_{A}^6-
\sum_{V}f_{V}^2 M_{V}^6\right]\log\frac{M_{Z}^2}{\mu^2}\,,
\eea where the first line is the result from the low--energy region and the
second
line the one from the high--energy region. We observe that the separation scale
$\mu^2$ cancels in the sum of the two terms; in other words, there is an exact
matching between the long--distance contribution and the short--distance
contribution. This cancellation also occurs in the LMD approximation to \QCD and
the result coincides then with the one given in eq.~(\ref{eq:CSMLMD}). The
reason
why we can exhibit this exact matching is of course due to the fact that the
function $\Pi_{LR}(Q^2)$ is an order parameter for all values of $Q^2$ and
therefore has duality properties under the transformation $Q^2\rightleftharpoons
{1}/{Q^2}$ which, in the large--$N_c$ limit, we have been able to work out
explicitly.

It is also interesting to look at eq.~(\ref{eq:OPEsplit}) from the point of
view of
effective field theory. The full integral in eq.~(\ref{eq:CZ}) can be split as
follows
\be\label{eq:effft} C_{Z}=\lim_{\mu\ra\infty}\left\{C_{Z}(M_{Z}\ra\infty,\mu)+
\delta C_{Z}(M_{Z},\mu)\right\}\, ,
\ee where $\delta C_{Z}(M_{Z},\mu) \equiv C_{Z}-C_{Z}(M_{Z}\ra\infty,\mu)$.
Each of
the two terms in the sum is now UV--divergent and this is why an
ultraviolet cutoff
$\mu$ is needed. The first term in eq.~(\ref{eq:effft}) corresponds to the
result
from the effective low--energy theory, where the $Z$ field has been
integrated out
and it is precisely given by the first line in the r.h.s. of
eq.~(\ref{eq:OPEsplit}). The second term $\delta C_{Z}(M_{Z},\mu)$
corresponds  to
the matching condition in the effective field theory language and the result is
precisely the one given in the second line of eq.~(\ref{eq:OPEsplit}).
Formally one
can set $\mu=M_{Z}$ (the effective field theory scale in this case) in these
expressions since they are actually
$\mu$ independent. Then $\delta C_{Z}(M_{Z},M_{Z}) =0$ and
$C_{Z}=C_{Z}(M_{Z}\ra\infty,M_{Z})$.

We are now in a position to discuss how the above calculation  proceeds in the
framework which is usually adopted to tackle weak matrix element
calculations.~\footnote{See e.g. the lectures of A.~Buras in ref.~\cite{Buras98}
and references therein.}

\section{Four--Quark Operators}
\label{sec:fourqo}

The relevant term in the Lagrangian of the Standard Model which is
responsible for the $Z$--induced contribution to the
$\pi^{+}-\pi^{0}$ mass difference is the neutral current interaction
term
\be
\cL_{\NC}=\frac{e}{2\sin\theta_{W}\cos\theta_{W}}
\left[\bar{q}_{L}\gamma^{\mu}T_{3}q_{L}-2\sin^{2}\theta_{W}
\bar{q}_{L}\gamma^{\mu}Q_{L}q_{L}-2\sin^{2}\theta_{W}
\bar{q}_{R}\gamma^{\mu}Q_{R}q_{R}\right]Z_{\mu}\,.
\ee When looking for the induced effective Lagrangian of order
$\cO(p^0)$ which contributes to Goldstone boson masses, it is
sufficient to consider left--right operators. In the absence of the
strong interactions, the effective four--quark Hamiltonian which
emerges after integrating out the $Z$ field  is
\bea\label{eq:eff}
 - \cH_{\eff} & = & \frac{-1}{M_{Z}^2}\frac{e^2}{\cos^2 \theta_{W}}\left[
\sin^2 \theta_{W} Q_{LR}
-\frac{1}{2}\left(\bar{q}_{L}\gamma_{\mu}T_{3}q_{L}\right)
\left(\bar{q}_{R}\gamma^{\mu}Q_{R}q_{R}\right)\right] \nonumber\\
 & = &
\frac{e^2}{M_{Z}^2} Q_{LR}-
\frac{e^2}{M_{Z}^2}\frac{1}{\cos^2
\theta_{W}}\frac{1}{6}\left(\sum_{q}\bar{q}\gamma_{\mu}q\right)
\left(\bar{q}_{R}\gamma^{\mu}Q_{R}q_{R}\right)\,,
\eea where
\be
 Q_{LR}\equiv\left(\bar{q}_{L}\gamma_{\mu}Q_{L}q_{L}\right)
\left(\bar{q}_{R}\gamma^{\mu}Q_{R}q_{R}\right)\,,
\ee and summation over quark colour indices within brackets is
understood. In fact, to $\cO(p^0)$,
only the first term proportional to the four--quark operator $Q_{LR}$
can contribute. In the presence of the strong interactions, the
evolution of $Q_{LR}$ from the scale
$M_{Z}^2$ down to a scale
$\mu^2$ can be calculated in the usual way, provided this $\mu^2$ is 
still large enough for a perturbative QCD (pQCD) evaluation to be valid. 
In the leading logarithmic approximation in
pQCD, and to leading non--trivial order in the $1/N_c$ expansion, the relevant
mixing in this evolution which we need to retain is simply given by
\be\label{eq:mixing}
Q_{LR}(M_{Z}^2) =
Q_{LR}(\mu^2)-3\frac{\als}{\pi}\frac{1}{2}\log\frac{M_{Z}^2}{\mu^2}
D_{RL}(\mu^2)+\cdots\,,
\ee where $D_{RL}$ denotes the four--quark density--density operator
\be D_{RL} \equiv
\sum_{q,q'} e_{q}e_{q'}(\bar{q'}_{L}q_{R})(\bar{q}_{R}q'_{L})\,,
\ee with $e_{q}$ and $e_{q'}$ the quark charges in units of the electric
charge. This can be seen as follows: in the $\overline{MS}$ renormalization
scheme,
the full evolution of the   Wilson coefficients $c_{Q}$ and $c_{D}$ of the
operators
$Q_{LR}$ and
$D_{LR}$ at the one loop level is governed by the equations 
(subleading contributions in the $1/N_c$ expansion have been neglected)
\be
\mu^2\frac{d}{d\mu^2}\left( \begin{array}{c} c_{Q} \\ c_{D} \end{array}\right)
=\frac{1}{4}\frac{\als N_c}{\pi}\left(\begin{array}{cc} \cdots & \cdots \\
\frac{6}{N_c} & -3+\cdots \end{array}\right)\left( \begin{array}{c}
c_{Q} \\ c_{D} \end{array}\right)\,,
\ee
with boundary conditions: $c_{Q}(M_Z)=1$ and
$c_{D}(M_Z)=\frac{3}{2}\frac{\als}{\pi}$. The result in eq.~(\ref{eq:mixing})
follows when taking $c_{D}(M_Z)=0$, which is appropriate when keeping 
the one-loop leading log only, and from the off--diagonal term in the 
(transposed) anomalous dimension matrix.

We are then confronted with a typical problem of bosonization of four--quark
operators.
The bosonization of $D_{RL}$ is only needed to leading order in the
$1/N_c$ expansion. To that order and to order $\cO(p^0)$ in the
chiral expansion it can be readily obtained from the bosonization of
the factorized density currents, with the result~\footnote{See e.g. the
lectures in
ref.~\cite{deR95} and references therein.}
\be
D_{RL} \equiv \sum_{q,q'}
e_{q}e_{q'}(\bar{q'}_{L}q_{R})(\bar{q}_{R}q'_{L})\ra
2B\frac{f_{\pi}^2}{4}\times 2B\frac{f_{\pi}^2}{4}
\tr \left(UQ_{L}U^{\dagger}Q_{R}\right)\,,
\ee
where $B$ is the low energy constant which describes the bilinear quark
condensate in the chiral limit,
\be
B=-\frac{\langle{\bar\psi}\psi\rangle}{f_{\pi}^{2}}\,.
\ee
We find that the overall contribution of the term proportional to the
$D_{RL}(\mu^2)$ four--quark operator, which we denote $C_{Z}\vert_{D_{RL}}$, is
given by the expression
\be\label{eq:4sd}
C_{Z}\vert_{D_{RL}} =\frac{-1}{M_{Z}^2}
3\frac{\als}{\pi}\frac{1}{4}\langle\bar{\psi}\psi\rangle^2
\frac{1}{2}\log\frac{M_{Z}^2}{\mu^2}\,,
\ee
and it is exactly the same result as the
one coming from the $d=6$ term of the OPE in the previous calculation,
the second line in eq.~(\ref{eq:OPEsplit}), once the duality constraint in
eq.~(\ref{eq:phi3}) is taken into account, {\it  i.e.} precisely the same
contribution as required
by the matching condition in the effective field theory analysis
above. Equation (\ref{eq:4sd}) does not change in an $\overline{MS}$
renormalization scheme when restricted to the one-loop leading log.

The problem is then reduced to the bosonization of the operator
$Q_{LR}(\mu^2)$. We are confronted here with a typical calculation of a 
hadronic matrix element of a four--quark operator, in this case the matrix
element
$\langle\pi^{+}\vert Q_{LR}(\mu^2)\vert\pi^{+}\rangle$. The factorized 
component
of the operator $Q_{LR}$, which is leading in $1/N_c$, cannot contribute to the
$\cO(p^0)$ term of the low--energy effective lagrangian. The contribution
we want
from this matrix element is therefore the next--to--leading one in
the $1/N_c$ expansion and it requires the evaluation of
non--factorizable four--quark matrix elements which is {\it a priori} a highly
non--trivial task. Yet, we know that there is a straightforward integral
representation of the observable we want to compute in terms of a two--point
function, and we have succeeded in doing the calculation that way within the
$1/N_c$--expansion framework. This seems to indicate that there should
be a way to do
the calculation of the matrix element $\langle\pi^{+}\vert
Q_{LR}(\mu^2)\vert\pi^{+}\rangle$ as well.
Indeed, we shall next show
that, for this particular matrix element, our present knowledge of analytic
results in non--perturbative QCD allows us to do the calculation exactly to
next--to--leading order in the $1/N_c$ expansion. The deep reason behind it is
that we know from the discussions in the previous sections that only two--point
functions can appear in the dynamics of this matrix element, and therefore it
is enough to have a formulation of the low--energy effective
lagrangian of \QCD compatible with the OPE constraints at short distances
for
two--point functions. These are precisely the constraints  which have
been recently discussed in ref.~\cite{KdeR97}.

The calculation proceeds along much the same lines as first
suggested in papers by Bardeen, Buras and G{\'e}rard~\cite{BBG87,Bardeen89}
sometime ago~\footnote{See also  refs.~\cite{BGK91,FG95,Ko98} and references
therein for more recent work.}, except that we shall go beyond loops
generated by Goldstone particle interactions alone in order to achieve a
correct matching with the logarithmic scale dependence of the
short--distance contribution in eq.~(\ref{eq:4sd}). We have evaluated the matrix
element
$\langle\pi^{+}\vert Q_{LR}(\mu^2)\vert\pi^{+}\rangle$
within the framework of an effective Lagrangian which is a
straightforward generalization to an arbitrary number of massive $J^P=1^-$
and $J^P=1^+$ mesonic states of the effective Lagrangian corresponding to
the LMD
approximation to \QCD recently discussed in ref.~\cite{PPdeR98}. Keeping only
the terms that are relevant for the evaluation of the $\cO(p^0)$
contributions, we
have
\bea
{\cal L}_{{\mbox{\rm {\scriptsize eff}}}} &=& \frac{f_\pi^2}{4}
\,\tr \,D_{\mu}U^+D^{\mu}U + L_{10}\,\tr\, U^+F_R^{\mu\nu}UF_{L,\mu\nu}
\nonumber\\
&&\ -\frac{1}{4}\sum_V \tr\left(
V_{\mu\nu}V^{\mu\nu}-2M_V^2V_{\mu}V^{\mu}\right)
-\frac{1}{4}\sum_A \tr\left( A_{\mu\nu}A^{\mu\nu}-2M_A^2A_{\mu}A^{\mu}\right)
\nonumber\\
&&\ -\frac{1}{2\sqrt{2}}\sum_V f_V\tr V_{\mu\nu}f_+^{\mu\nu}
-\frac{1}{2\sqrt{2}}\sum_A f_A\tr A_{\mu\nu}f_-^{\mu\nu}
+\cdots\,,
\eea
with the same notation as in ref.~\cite{PPdeR98},
$L_{10}$ being one of the $\cO(p^4)$ constants of the Gasser-
Leutwyler Lagrangian~\cite{GL85}.

The bosonized expressions of the currents
$\left(\bar{q}_{L}\gamma^{\mu}Q_{L}q_{L}\right)$ and
$\left(\bar{q}_{R}\gamma^{\mu}Q_{R}q_{R}\right)$ are obtained from the
resulting effective action
$\Gamma_{{\mbox{\rm {\scriptsize eff}}}}$ at tree level,
\be
\left(\bar{q}_{L}\gamma^{\mu}Q_{L}q_{L}\right) \ra
(Q_{L})_{ij}\frac{\delta\Gamma_{{\mbox{\rm {\scriptsize eff}}}}}
{\delta l_{\mu}^{ij}(x)}\quad\annd\quad
\left(\bar{q}_{R}\gamma^{\mu}Q_{R}q_{R}\right) \ra
(Q_{R})_{ij}\frac{\delta\Gamma_{{\mbox{\rm {\scriptsize eff}}}}}
{\delta r_{\mu}^{ij}(x)}\,,
\ee
where $i,j=u,d,s$ are flavour indices. 
The contributions from a loop of virtual pions, vector and axial--vector
resonances
are computed by applying a finite ultraviolet cut-off $\mu^2$ 
to the integral over the
corresponding (euclidean) virtual loop momentum.
One technical point, as already
discussed in ref~\cite{PdeR91} (see eqs.~(7.24) to
(7.26) of this reference), requires however attention. It is the fact that
to next--to--leading order  in
the $1/N_c$ expansion, and besides the loops generated by the bosonized factorized
currents that we have just discussed, there
are also contact terms generated by the second variation of the effective
action. In our case these terms are:
\bea
Q_{LR}^{{\mbox{\rm {\scriptsize contact}}}} & =
& (Q_{L})_{ij}(Q_{R})_{kl}\frac{\delta^2\Gamma_{{\mbox{\rm {\scriptsize
eff}}}}}{\delta l_{\mu}^{ij}(x)\delta r^{\mu, kl}(x)} \nn
\\
 & = & \left\{-2f_{\pi}^2\,\tr\left( Q_{R}UQ_{L}U^{\dagger}\right)\delta(0)-
6L_{10}\,\tr\left(Q_{R}UQ_{L}U^{\dagger}\right)\Box\delta
(0)+\cdots\right\}\,,
\eea
where the $\delta$--function contributions are to be interpreted within the
same cut-off regularization~:
\be
\delta(0)\ra \pi\int_{0}^{\mu^2} dQ^2 Q^2  = \frac{\pi}{2}\mu^4
\qquad\annd\qquad
\Box\delta(0) \ra \pi\int_{0}^{\mu^2} dQ^2 Q^4  = \frac{\pi}{3}\mu^6\,.
\ee
Adding the two types of contributions, the final result reads
\bea\label{eq:QLRresult}
\langle\pi^{+}\vert
Q_{LR}(\mu^2)\vert\pi^{+}\rangle\ &  = & \ \frac{+1}{16\pi^2}
\left\{ \frac{3}{2}\mu^4 + 4\frac{L_{10}}{f_{\pi}^2}\mu^6 + \right. \nn \\
 &  & \left. +\frac{3}{f_{\pi}^2}\int_{0}^{\mu^2} dQ^2 Q^6
\bigg[ \sum_{V}\frac{f_{V}^2}{Q^2+M_V^2} -
       \sum_{A}\frac{f_{A}^2}{Q^2+M_A^2}\bigg]\right\}\,.
\eea
The contribution of the Goldstone bosons alone corresponds to
the two terms in the first line of the r.h.s. of eq.~(\ref{eq:QLRresult}). They
display a typical polynomial
dependence with respect to the cut-off $\mu$, which can hardly provide a
reasonably good matching with the logarithmic scale dependence coming from
the short--distance contributions in
$C_{Z}\vert_{D_{RL}}$. In fact, in an $\overline{MS}$ regularization scheme, as
commonly chosen for the evaluation of the short--distance Wilson coefficients,
these power divergences will automatically disappear.  Simply adding higher
resonances does not by itself solve the problem of matching the long-- and the
short--distances either; however, when the information coming from the
short--distance properties of the correlator in eq.~(\ref{eq:lrtpf}), {\it i.e.}
the two Weinberg sum rules (\ref{eq:weinbergsrs}) and the sum
rule~\cite{GL84,KdeR97}
\be\label{eq:L10}
-4L_{10}=\sum_{V} f_{V}^2-\sum_{A}f_{A}^2\,,
\ee
is taken into account, the result of eq.~(\ref{eq:QLRresult}) can indeed be
recast
into a form which reproduces the first line of eq.~(\ref{eq:OPEsplit}). Notice
that in an $\overline{MS}$ regularization scheme, the integral in eq.
(\ref{eq:QLRresult}) should be understood in $n=4-\epsilon$ dimensions and therefore
multiplied by
$\mu^{4-n}$ with $\mu$ the
$\overline{MS}$ regularization scale. The result, when combined with
eq.~(\ref{eq:4sd}) finally yields eq.~(\ref{eq:OPEsplit}) once again. We
insist on
the fact that, regardless of the regularization one chooses,  
the calculation we
have done of the matrix element $\langle\pi^{+}\vert
Q_{LR}(\mu^2)\vert\pi^{+}\rangle$ is an exact calculation to next--to--leading
order in the $1/N_c$ expansion and to $\cO(p^0)$ in the chiral expansion.

\section{Comments and Outlook}
\label{sec:conlook}

The main purpose of the above analysis has been to show that the
calculation of the
electroweak contribution to the $\pi^{+}-\pi^{0}$ mass difference in the chiral
limit is an interesting theoretical 
laboratory for testing issues connected with
the matching of long and short distances in the evaluation of hadronic weak
matrix elements in general.

On the one hand, upon identifying the relevant low--energy constant in
terms of the appropriate QCD correlator, we have been able to obtain, in the
large--$N_c$ limit, an exact result in terms of resonance parameters
with masses and couplings constrained by the short--distance
properties of this correlator.

On the other hand, due to the presence of the large scale set by $M_Z$, we
were able to proceed as done usually in the study of weak non--leptonic
processes.
In this respect, we encounter a typical situation: the effective
Hamiltonian at tree level is given by a current--current four quark
operator $Q_{LR}$, which, upon taking into account short--distance pQCD
corrections, mixes with the density--density operator $D_{RL}$. The matrix
element of the latter to $\cO(p^0)$, 
which is only needed at the leading order in the large--$N_c$
limit, is easily computed. In the same limit, $\langle\pi^{+}\vert
Q_{LR}(\mu^2)\vert\pi^{+}\rangle$ vanishes, and next--to--leading,
non--factorizable, contributions have to be included.
A suitable matching of the long--distance contribution with the logarithmic
scale dependence coming from the short--distance contribution cannot be
achieved, in this case, by evaluating the relevant hadronic matrix elements
in terms of an effective theory incorporating only the Goldstone bosons.
When the
contribution of vector and axial--vector states satisfying the appropriate
short--distance constraints are included, we find an exact matching.

We think that the observations which follow from this example point towards
promising perspectives for a systematic determination of hadronic weak  matrix
elements within the framework of \QCD. The required steps are the
following: first
one needs an identification of the low--energy constants of the $\Delta S=1$ and
$\Delta S=2$ effective chiral Lagrangian in terms of integrals of QCD Green's
functions, {\it i.e.} the equivalent of our eq.~(\ref{eq:CSM}) above. Next one
should
proceed to the study of the constraints which the OPE imposes on these Green's
functions in the large--$N_c$ limit, {\it i.e.} the equivalent of the Weinberg
sum rules
in eqs.~(\ref{eq:weinbergsrs}) and the sum rules in eqs.~(\ref{eq:phi3}) and
(\ref{eq:L10}). The final step is the construction of an effective
Lagrangian, along 
the lines shown in ref.~\cite{PPdeR98}, which incorporates these
constraints. This
program is at present under study.

In the meantime, it would be interesting to see
a calculation done using numerical simulations of lattice QCD of the matrix
element
$\langle\pi^{+}\vert Q_{LR}(\mu^2)\vert\pi^{+}\rangle$, so as to be able to
appreciate the performance of the lattice approach in a simple case which we
understand well analytically.

\vspace*{7mm} {\large{\bf Acknowledgments}}
\vspace*{3 mm}

\noi
We wish to thank J.~Bijnens, M.~Perrottet, A.~Pich and J.~Prades for
discussions on topics related to the work reported here.  This work has been
supported in part by TMR, EC-Contract No. ERBFMRX-CT980169 (EURODA$\phi$NE). The
work of S.~Peris has also been partially supported by the research project
CICYT-AEN95-0882.


\end{document}